\documentclass[epj]{svjour}

\usepackage{amsfonts}
\usepackage{amsmath}
\usepackage{graphicx}
\usepackage{latexsym}

\newcommand{\supp}{\operatorname{supp}}
\newcommand{\Esf}{\mathsf{E}}

\begin{document}

\title{`Measurement of Quantum Mechanical Operators' Revisited}
\author{Leon Loveridge\thanks{\email{ldl500@york.ac.uk}} \and Paul Busch\thanks{\email{paul.busch@york.ac.uk}}}

\institute{University of York, Department of Mathematics, Mathematical Physics Section,
York, UK}

\date{{\bf To appear in: Eur.~Phys.~J.~D (2011)}}

\abstract{
The Wigner-Araki-Yanase (WAY) theorem states a remarkable
limitation to quantum mechanical measurements in the presence of 
additive conserved quantities.
Discovered by Wigner in 1952, this limitation is known to induce constraints on the
control of individual quantum systems in the
context of information processing. It is therefore important to understand
the precise conditions and scope of the WAY theorem. Here we elucidate its
crucial assumptions, briefly review some generalizations, and show how a
particular extension can be obtained by a simple modification of the
original proofs. We also describe the evolution of the WAY theorem from
a strict no-go verdict for certain, highly idealized, precise measurements
into a quantitative constraint on the accuracy and approximate
repeatability of imprecise measurements.
}

\maketitle

\section{Introduction}\label{sec:intro}

Quantum mechanical experiments involving the manipulation of individual
quantum objects no longer reside only in the minds of a few theoretical
physicists, but are a routine occurrence across many physical disciplines
such as quantum optics and quantum information. This not only provides new
and exciting opportunities for future technologies such as quantum
computing, but necessitates a fundamental re-examination of the quantum
mechanical formalism itself, and a new understanding of its role in modern
applications. With the ever-decreasing size of the components involved in
these technologies, it is both interesting from a foundational viewpoint and
important in more practical respects to understand any fundamental
limitations on the possible size of such microscopic instruments. 

One such limitation arises as a consequence of conservation laws for 
additive quantities that do not commute with the observable to be measured. 
Whilst considering spin-$\frac 12$ measurements,
Wigner \cite{wigner} discovered that the total angular momentum of the object plus apparatus
cannot be conserved in an accurate and repeatable measurement of a particular component.
This observation was soon stated 
in greater generality as a theorem by Araki and Yanase \cite{way} 
that has become known as the Wigner-Araki-Yanase (WAY) theorem. Despite
the fact that the original papers \cite{wigner} and \cite{way} have been widely
noted and the WAY theorem has been extended in various respects, its full
scope is still unknown. 

It is the purpose of this paper to survey the evolution of formulations of WAY-type 
theorems, % (sec.~\ref{sec:overview}) 
elucidate the significance of the underlying assumptions, and clarify the general 
structure and extent of such theorems. We will also provide some new extensions
of known results and propose an answer to a long-standing question
concerning the possibility of momentum-conserving measurements of the position
of a quantum particle.

In Sec.~\ref{sec:wigner} we revisit Wigner's 1952 paper \cite{wigner}. 
In particular 
we scrutinize the final page where Wigner examines the consequences of dropping the assumption 
that the measurement be repeatable. This is a relaxation which is physically relevant, but is still 
not appreciated by many practitioners of quantum theory. Wigner notes that in this case the issue 
arises of the distinguishability of the states of the measuring apparatus, given that the limitation 
imposed by the conservation law also applies to a measurement of the pointer. The paper \cite{wigner}
is written (in German) with the simplicity and elegance characteristic of Wigner; in order to
make it more widely accessible, a translation into English is provided as a concurrent 
publication \cite{wigner-en}. 

In Sec. \ref{sec:ay} we proceed to give a modification of the proof of Araki and Yanase \cite{way} 
leading to a sharpening and extension of the WAY theorem. They prove for certain classes of 
observables and conserved quantities that under the assumption of accuracy and repeatability, 
the observable to be measured must commute with the (object part of) the conserved quantity. 
Here we show that the same conclusion follows if the repeatability of the measurement is 
replaced by the assumption that the pointer observable commutes with the conserved quantity. 
This condition, which following Ozawa \cite{ozawacon} we shall call \emph{Yanase condition}, 
was already alluded to in \cite{wigner} and  \cite{yanase-opt}. In fact, the WAY theorem also 
precludes accurate and repeatable measurements of the pointer observable (given the conservation law)
unless the Yanase condition is fulfilled.

In Sec. \ref{sec:approx} we review formulations of WAY-type limitations for approximate measurements. 
In particular we present and develop an inequality first formulated by Ozawa \cite{ozawacon} that demonstrates
trade-off relations between a measure of error and the ``size" of the apparatus 
(suitably defined). In Sec.~\ref{sec:wayout}  we revisit some model measurement schemes, notably by 
Ohira and Pearle \cite{ohira/pearle}, and observe that the ``ways out" of the WAY limitation sought there 
always come at the expense of violating the repeatability \emph{and} Yanase conditions. This
helps to highlight the fact that the WAY theorem is often paraphrased in a superficial way, ignoring the
repeatability property and  the relevance of the Yanase condition.

Sec. \ref{sec:pos} contains a description of the largely unexplored question of 
whether position measurements that respect momentum conservation are subject to a WAY-type 
limitation. Here we adapt Ozawa's inequality to establish the necessity
of a large apparatus for good measurements, provided that 
the Yanase condition is satisfied. We also formulate a trade-off inequality analogous to Ozawa's 
inequality with which one can quantify the degree of repeatability achievable given the size of the 
apparatus. Finally we provide an affirmative answer, in a certain approximate sense, to a problem 
posed by Stein and Shimony in 1979 \cite{Stein-Shimony} concerning the feasibility of repeatable
position measurements obeying momentum conservation.

The paper concludes with some remarks on the relavance of the WAY theorem in contemporary quantum physics and
quantum information.

We begin with an outline of concepts of quantum measurement theory relevant to our investigation.

\section{Preliminaries}\label{sec:prelim}

We will apply the standard formulation of quantum measurements (e.g., \cite{QTM})
where the quantum system and apparatus are represented by complex Hilbert spaces $\mathcal{H}$ 
and $\mathcal{K}$, respectively. These come equipped with the usual inner products denoted
$\left\langle \cdot|\cdot\right\rangle$. Observables are given as positive operator
valued measures (POVMs) $\mathsf{E}:X\mapsto \mathsf{E}(X)$. The operators $\mathsf{E}(X)$ associated with subsets 
$X$ of $\mathbb R$ are positive operators whose expectation values $\langle\psi |\Esf(X) \psi \rangle$ (for
normalized states $\psi\in\mathcal{H}$) represent the measurement outcome probabilities of finding 
a result in $X$; these operators are called {\em effects}. If these effects are projections, the POVM $\mathsf{E}$
is a spectral measure from which one recovers the standard representation of an observable as a self-adjoint
operator, namely $\int x\mathsf{E}(dx)\equiv M$. 

For a measurement to take place, there must 
be an interaction between the object system and a macroscopic measuring
apparatus, whereby the experimenter can read off the values of the measured observable. The part of 
this device that actually comes into contact with the quantum system under investigation may only be a 
small component of the whole apparatus (and could be referred to as a {\em probe}). We shall not 
discuss the process of amplification by which information from the interaction generates macroscopic 
outcome values. 

The composite system-apparatus Hilbert space is described by the tensor product 
$\mathcal{H}_{T}:=\mathcal{H}\otimes \mathcal{K}$. The time evolution of the system-plus-apparatus is then given by a unitary
operator $U$ on $\mathcal{H}_{T}$, which serves to couple the states of the
system to those of the apparatus during an interaction period $\tau $. In order
that this interaction may be said to lead to a measurement of an observable $\Esf$, some extra
elements are still required; these are a self-adjoint operator $Z$ on $%
\mathcal{K}$, which represents a \textquotedblleft pointer
observable\textquotedblright , and a (fixed) initial apparatus state in $%
\mathcal{K}$, chosen to be the pure (vector) state $\phi $, with $\left\Vert
\phi \right\Vert=1$.

We then define a measurement ${\mathcal{M}}$ of an observable $\Esf$
as the 5-tuple $\langle \mathcal{K},U,\phi ,Z,f\rangle $ satisfying the \emph{probability 
reproducibility condition}; that the outcome
distribution for $\Esf$ in any state $\varphi $ be recovered from the pointer
statistics in the final state $\Psi _{\tau }=U(\varphi \otimes \phi )\in 
\mathcal{H}_{T}$. This condition can be written
symbolically as 
\begin{equation}
\left\langle \Psi _{\tau }|\boldsymbol{1}\otimes \Esf^{Z}(f^{-1}(X))\Psi _{\tau}
\right\rangle \equiv \left\langle \varphi |\Esf(X)\varphi \right\rangle,
\label{PRC}
\end{equation}
where $ \Esf^{Z}(f^{-1}(X))$ are spectral projections of $Z$, and \eqref{PRC} holds for all $\varphi$ and $X$. Conversely, given a
measurement scheme as described above, this relation determines the measured
observable $\Esf$. The scaling function $f$ is used to map the values of the pointer
to those of the measured observable.

A measurement is said to be \emph{repeatable }if, upon immediate repetition
of the measurement, the same outcome is achieved with certainty. This may be written:
\begin{equation}
\left\langle \Psi _{\tau} |\Esf(X)\otimes \Esf^{Z}(f^{-1}(X))\Psi _{\tau}
\right\rangle =\left\langle \varphi |\Esf(X)\varphi \right\rangle.
\end{equation} \label{rep}
It should be noted that even when \eqref{PRC} holds, it is  not guaranteed 
that \eqref{rep} is satisfied, and as such the question of repeatability must be treated independently 
of that of probability reproducibility. Conditions.~\eqref{PRC} and \eqref{rep} can be rephrased in a more
concise form as follows: if the outcome of the first measurement is described
by a set $X$, the system's state is obtained by taking a partial trace operation: 
\begin{equation}
\rho_X={\rm tr}_{\mathcal K}[\boldsymbol{1}\otimes \Esf^{Z}(f^{-1}(X))|\Psi _{\tau}\rangle\langle \Psi _{\tau}|];
\end{equation}
Now condition \eqref{PRC} reads:
 \begin{equation}
 {\rm tr}[\rho_X]=\left\langle \varphi |\Esf(X)\varphi \right\rangle.
 \end{equation} 
Writing $\hat\rho_X=\rho_X/{\rm tr}[\rho_X]$ for the normalized state, the  repeatability condition \eqref{rep} 
then becomes:
\begin{equation}
{\rm tr}[\hat\rho_X\,\Esf(X)]=1.
\end{equation}

Although in many textbooks the term \emph{measurement} is understood as comprising
the repeatability property, it is important to recognize that most realistic measurements
are not repeatable. Furthermore, measurements are rarely accurate and the actually measured
observable is appropriately described as a POVM. We will see below that even as early as 1952 Wigner was
working with more general measurement models that do not satisfy the
repeatability criterion and whose associated observable is a POVM.

\section{Wigner 1952}\label{sec:wigner}

\subsection{Wigner's example}\label{subsec:wigner-example}

Wigner first noticed that repeatable measurements of the 
$x$-component of the spin of a spin-$\frac{1}{2}$ particle 
necessarily violate the conservation of the $z$-component of the total
angular momentum of the system plus apparatus, written $S_{z}\otimes \boldsymbol{1}+
\boldsymbol{1}\otimes J_{z}$. He also demonstrated the feasibility
of recovering arbitrarily accurate and repeatable measurements if the apparatus
becomes ``large''. This is a significant feature in much of the work following
Wigner's discovery, and we sketch the argument here. 
We continue with the notation that $\phi\in\mathcal{K}$ represents the initial (normalized)
apparatus state, and $\phi _{\pm }\in\mathcal{K}$ orthonormal pointer states, and throughout we shall choose
units where $\hbar =1$. The unitary evolution %on $\mathcal{H}\otimes\mathcal{K}$ 
takes the form (with $\varphi _{\pm }$ representing
$S_{x}$ eigenstates):
\begin{eqnarray}
\varphi _{+}\otimes \phi &\longrightarrow &\varphi _{+}\otimes \phi _{+},
\label{intro1} \\
\varphi _{-}\otimes \phi &\longrightarrow &\varphi _{-}\otimes \phi _{-};
\label{intro2}
\end{eqnarray}%
 the evolution for the eigenstates 
$\psi _{\pm }=(\varphi _{+}\pm \varphi _{-})/ \sqrt{2}$ of $S_{z}$ is then
\begin{eqnarray}
\psi _{+}\otimes \phi &\longrightarrow &\frac{1}{2}\left[ \psi _{+}\otimes
(\phi _{+}+\phi _{-})+\psi _{-}\otimes (\phi _{+}-\phi _{-})\right] ,
\label{contra} \\
\psi _{-}\otimes \phi &\longrightarrow &\frac{1}{2}\left[ \psi _{+}\otimes
(\phi _{+}-\phi _{-})+\psi _{-}\otimes (\phi _{+}+\phi _{-})\right] .
\label{contra 2}
\end{eqnarray}%
This violates angular momentum conservation, since the expectations 
$\left\langle S_{z}+J_{z}\right\rangle $ agree on the right
hand sides of (\ref{contra}) and (\ref{contra 2}) but differ by one unit on
the left hand sides. Since, as Wigner argues, spin component measurements
are ``practically possible'', he introduces
the following modification in order to model an approximate realization of
the measurement:
\begin{eqnarray}
\varphi _{+}\otimes \phi &\longrightarrow &\varphi _{+}\otimes \phi
_{+}+\varphi _{-}\otimes \eta , \\
\varphi _{-}\otimes \phi &\longrightarrow &\varphi _{-}\otimes \phi
_{-}-\varphi _{+}\otimes \eta ,
\end{eqnarray}%
with $\left\langle \eta ,\phi _{\pm }\right\rangle =0$. There are now three
(un-normalized) pointer states,\ representing a three-outcome measurement,
the third (labelled by $\eta $) corresponding to an undetermined spin, representing
a situation where the apparatus cannot identify a definite spin. The
two definite outcomes are represented by effects $E_{\pm }=(1- \left\Vert
\eta \right\Vert ^2) P[\varphi_{\pm}]$, and the third is represented by a trivial effect  
$E_0=\left\Vert\eta \right\Vert ^2\mathbf{1}$ (with probability given by $\left\Vert\eta \right\Vert ^2 $).
Wigner shows that $\left\Vert\eta \right\Vert ^2$ can be made arbitrarily small
given a ``large'' apparatus. Specifically he shows that if
the state $\phi $ has a very large number of components in its expansion in terms of $J_z$-eigenvectors
$\phi_\nu$, so that $\phi =\sum\limits_{\nu =1}^{n}\phi _{\nu}$, then 
with some suitable assumptions and conditions, $\left\Vert\eta \right\Vert ^2=1/(2n-1)$.  
Thus in the large-$n$ limit, $\left\Vert
\eta \right\Vert \rightarrow 0$ and accurate and repeatable measurements
are, to a very good approximation, recovered. 

We note that the large size of the apparatus is used here only as a sufficient condition to achieve good
measurement accuracy; the argument does not yield it as a necessary one.

\subsection{Implications of dropping repeatability}\label{subsec:alternative}

Wigner's consideration in his final page is intriguing although very sketchy and 
somewhat open-ended; there he discusses a more general measurement scheme in
which the repeatability restriction is dropped. We carefully reconstruct his
argument in the Appendix;  here we provide a more concise and more
general calculation, which contains Wigner's conclusion as a special case. This approach
has considerably less cumbersome algebra, and  relies on exploiting the condition that the
interaction must be a measurement (in the sense of \eqref{PRC}) from the beginning.\ We make
no assumption on the product form of the final states, and allow the most
general (entangled) final state in the system-apparatus Hilbert space.

For notational convenience and following Wigner,
when required we shift the spectral values of the observables concerned in order that they are integers; for example
the eigenvalues of the object part of the conserved quantity are now $0$ and $1$. In contrast to Wigner, we do not make the assumption
that the spectrum of the apparatus' conserved quantity is bounded below by zero. With 
$\chi_{k}^{\prime}$, $\chi_{k}^{\prime \prime}$, $\phi_{k}^{\prime}$ and $\phi_{k}^{\prime \prime}$ 
representing (un-normalized) eigenstates of $J_{z}$ and $\psi _{0}$, $\psi _{1}$ (normalized) $S_{z}$ eigenstates,
the unitary evolution $U$ gives:
\begin{eqnarray}
(\psi _{0}+\psi _{1})\otimes \sum \phi _{k} &\overset{U}{\longrightarrow } 
&\psi_{0}\otimes \sum \phi _{k}^{\prime }+\psi _{1}\otimes \sum \chi _{k}^{\prime},  \qquad \label{wiggen1} \\
(\psi _{0}-\psi _{1})\otimes \sum \phi _{k} &\overset{U}{\longrightarrow } 
&\psi_{0}\otimes \sum \phi _{k}^{\prime \prime }+\psi _{1}\otimes \sum \chi_{k}^{\prime \prime } . \label{wiggen2}
\end{eqnarray}
In order to exploit the conservation law we
take sums and differences of (\ref{wiggen1}) and (\ref{wiggen2}), and obtain 
\begin{align}
2\psi _{0}\otimes \sum \phi _{k}\longrightarrow\psi _{0}\otimes &\sum
(\phi _{k}^{\prime }+\phi _{k}^{\prime \prime }) \nonumber\\
&+\psi _{1}\otimes \sum (\chi_{k}^{\prime }+\chi _{k}^{\prime \prime }) , \\
%\end{multline}
%\begin{multline}
2\psi _{1}\otimes \sum \phi _{k}\longrightarrow\psi _{0}\otimes &\sum
(\phi _{k}^{\prime }-\phi _{k}^{\prime \prime }) \nonumber \\
&+\psi _{1}\otimes \sum (\chi_{k}^{\prime }-\chi _{k}^{\prime \prime}).
\end{align}%
The conservation law now entails  that for any $k$: 
\begin{eqnarray} 
\label{wigspec1}
2\psi _{0}\otimes \phi _{k} &\longrightarrow 
&\psi _{0}\otimes (\phi
_{k}^{\prime }+\phi _{k}^{\prime \prime })+\psi _{1}\otimes (\chi
_{k-1}^{\prime }+\chi _{k-1}^{\prime \prime }),\qquad \\
\label{wigspec2}
2\psi _{1}\otimes \phi _{k} &\longrightarrow 
&\psi _{0}\otimes (\phi
_{k+1}^{\prime }-\phi _{k+1}^{\prime \prime })+\psi _{1}\otimes (\chi
_{k}^{\prime }-\chi _{k}^{\prime \prime }).
\end{eqnarray}%
At this point we wish to make contact with Wigner's work, and so specify 
that the apparatus carries no units of the conserved quantity.
This is implemented by setting $k=0$, and so $\phi=\phi _{0}$. With this stipulation
and allowing
for the fact that, in general, the final apparatus states may have negative 
angular momentum values, we combine (\ref{wigspec1}) and (\ref{wigspec2}) to obtain:
\begin{align}
(\psi _{0}&+\psi _{1})\otimes \phi _{0} \longrightarrow \frac{1}{2}\psi
_{0}\otimes (\phi _{0}^{\prime }+\phi _{1}^{\prime }+\phi _{0}^{\prime
\prime }-\phi _{1}^{\prime \prime })\nonumber\\
&+\frac{1}{2}\psi _{1}\otimes (\chi
_{-1}^{\prime }+\chi _{-1}^{\prime \prime }+\chi _{0}^{\prime }-\chi
_{0}^{\prime \prime }), \\
(\psi _{0}&-\psi _{1})\otimes \phi _{0} \longrightarrow \frac{1}{2}\psi
_{0}\otimes (\phi _{0}^{\prime }-\phi _{1}^{\prime }+\phi _{0}^{\prime
\prime }+\phi _{1}^{\prime \prime })\nonumber\\
&+\frac{1}{2}\psi _{1}\otimes (\chi
_{-1}^{\prime }+\chi _{-1}^{\prime \prime }-\chi _{0}^{\prime }+\chi
_{0}^{\prime \prime }).
\end{align}%
From here it follows, by comparison with (\ref%
{wiggen1}) and (\ref{wiggen2}), that $\phi _{0}^{\prime \prime }=\phi
_{0}^{\prime }$, $\phi _{1}^{\prime \prime }=-\phi _{1}^{\prime }$, $\chi
_{-1}^{\prime \prime }=\chi _{-1}^{\prime }$, $\chi _{0}^{\prime \prime
}=-\chi _{0}^{\prime }.$ Thus%
\begin{align}
(\psi _{0}+\psi _{1})\otimes \phi _{0}&\longrightarrow \nonumber \\ 
&\psi _{0}\otimes (\phi_{0}^{\prime }+\phi _{1}^{\prime })
+\psi _{1}\otimes (\chi _{0}^{\prime}+\chi _{-1}^{\prime }),  \label{gen1}\\
%\end{multline}
%\begin{multline}
(\psi _{0}-\psi _{1})\otimes \phi _{0}&\longrightarrow \nonumber \\ 
&\psi _{0}\otimes (\phi_{0}^{\prime }-\phi _{1}^{\prime })
+\psi _{1}\otimes (-\chi _{0}^{\prime}+\chi _{-1}^{\prime }).  \label{gen2}
\end{align}%
Taking the partial trace over the system's Hilbert space in (\ref%
{gen1}) and (\ref{gen2}) yields (mixed) reduced probe states $\rho ^{+}$ and 
$\rho ^{-}$ respectively. With $\{e_i\}$ an arbitrary orthonormal basis in $\mathcal{K}$,
\begin{equation}
\rho ^{\pm }:=\mathrm{tr}_{\mathcal{H}}(P[U(\varphi ^{\pm }\otimes \phi _{0}
)])=\sum \left\langle e_{i}|P[U(\varphi ^{\pm }\otimes \phi
)e_{i}\right\rangle,  \label{reduced}
\end{equation}%
where $P[U(\varphi ^{\pm }\otimes \phi _{0} )]$ are the orthogonal projections
onto the final states, and $\varphi ^{\pm }=\frac{1}{\sqrt{2}}(\psi _{0}\pm
\psi _{1})$.\ For $U$ to yield a measurement in the sense of
\eqref{PRC}, it is
required that the reduced states corresponding to two orthogonal initial
states are unambiguously distinguishable; that is that they are supported on
orthogonal subspaces of $\mathcal{K}$. This is equivalent to the statement
that $\mathrm{tr}(\rho ^{+}\rho ^{-})$ must vanish, and it readily emerges
that%
\begin{multline}
0=\mathrm{tr}(\rho ^{+}\rho ^{-})=(\left\Vert \phi _{0}^{\prime }\right\Vert
^{2}-\left\Vert \phi _{1}^{\prime }\right\Vert ^{2})^{2}\\+(\left\Vert \chi
_{-1}^{\prime }\right\Vert ^{2}-\left\Vert \chi _{0}^{\prime }\right\Vert
^{2})^{2}+2\left\vert \left\langle \phi _{0}^{\prime }|\chi _{0}^{\prime
}\right\rangle \right\vert ^{2} . \label{trace}
\end{multline}%
Since each term in this sum is non-negative, it follows that they must each
vanish individually, and so $\left\Vert \phi _{0}^{\prime }\right\Vert
^{2}=\left\Vert \phi _{1}^{\prime }\right\Vert ^{2}$, $\left\Vert \chi
_{-1}^{\prime }\right\Vert ^{2}=\left\Vert \chi _{0}^{\prime }\right\Vert
^{2}$ and $\left\langle \phi _{0}^{\prime }|\chi _{0}^{\prime }\right\rangle
=0$. \ Hence (\ref{trace}) is only satisfied if either $%
\phi _{0}^{\prime }=\phi _{1}^{\prime }=0$ or $\chi _{-1}^{\prime }=\chi
_{0}^{\prime }=0$, since $\phi_{0}^{\prime }$ and $\chi_{0}^{\prime }$ are collinear.
\ There are two scenarios to consider: first, where 
$\chi _{-1}^{\prime }=\chi _{0}^{\prime }=0$ and the measurement takes the form 
\begin{align}\label{mmt1}
(\psi _{0}+\psi _{1})\otimes \phi _{0} \longrightarrow \psi _{0}^{\prime }\otimes
(\phi _{0}^{\prime }+\phi _{1}^{\prime }),\\
(\psi _{0}-\psi _{1})\otimes \phi _{0} \longrightarrow \psi _{0}^{\prime }\otimes
(\phi _{0}^{\prime }-\phi _{1}^{\prime })\label{mmt2} .
\end{align}
This is the form that Wigner arrives at on his final page (see our Appendix). 
The second scenario is given by $\phi _{0}^{\prime }=\phi _{1}^{\prime }=0$ where
\begin{align} \label{mmt3}
(\psi _{0}+\psi _{1})\otimes \phi _{0} \longrightarrow \psi _{1}^{\prime
}\otimes (\chi _{0}^{\prime }+\chi _{-1}^{\prime }),\\ 
(\psi _{0}-\psi _{1})\otimes \phi _{0} \longrightarrow \psi _{1}^{\prime
}\otimes (-\chi _{0}^{\prime }+\chi _{-1}^{\prime })\label{mmt4}  .
\end{align}

It is now easy to verify the unitarity of the interaction. The measurement property 
guarantees that $\phi _{0}^{\prime }$ and $\phi _{1}^{\prime }$ have
equal (squared) norm, as do $\chi _{0}^{\prime }$ and $\chi _{-1}^{\prime }$, leaving the right 
hand sides of \eqref{mmt1} and \eqref{mmt2} orthogonal, and so too \eqref{mmt3} and \eqref{mmt4}. 
For both scenarios, the final system state is independent of the initial one,
and repeatability is clearly violated. 

It seems that dropping the requirement of repeatability
has allowed for the possibility of an accurate measurement, whereas before
this was ruled out by the non-commutativity of $S_{x}$ with $J_{z}$.
Furthermore, here Wigner has chosen $\phi$ to be an
eigenstate of the conserved quantity with eigenvalue zero, whereas we
saw in the previous subsection  that he chose $\phi$ to have
very many components in order to overcome the limitation imposed by the 
conservation law. Hence giving up repeatability also seems to take away the
 size constraint for the apparatus.

However, Wigner points out (and this has also been noted in \cite{yanase-opt}) that
the issue of a measurement limitation due to the conservation law has been transferred 
from the system to the apparatus, since (as is made evident above) the final apparatus states
must be eigenstates of the $x$-component of the apparatus' angular momentum 
yielding a pointer observable that does not commute with $J_{z}$. It is natural to
consider a pointer reading to be an instance of a repeatable measurement, since otherwise
there would be no stable record of the measurement (see also \cite{ozawacon}).
Here the WAY-type limitation reappears at the level of the pointer observable, which turns out not
to commute with the apparatus' conserved quantity. Hence the Yanase condition appears to be
violated necessarily. Wigner, it seems, was moving toward a general no-go result: that
if one wishes to have an accurate measurement, both repeatability and the
Yanase condition must be abandoned. Indeed this is the case, as shall be
proved in the next section. %Sec.~\ref{sec:ay}.

\section{The WAY Theorem}\label{sec:ay}
\subsection{The work of Araki and Yanase extended}\label{subsec:A+Y}

Araki and Yanase \cite{way} took up the work of Wigner and proved a general
theorem which we state and prove in a somewhat extended and sharpened form. 
We show that for the same conclusion to be drawn
the assumption of repeatability can be replaced by  the Yanase condition.  
 
Let  $L=L_{1}\otimes \boldsymbol{1}+\boldsymbol{1}\otimes L_{2}$ denote the conserved quantity
and $M$ the operator we wish to measure.

\noindent \textbf{Theorem}
\emph{Let ${\mathcal{M}}:=\langle \mathcal{K},U,\phi ,Z,f\rangle $ be a 
measurement of a discrete-spectrum self-adjoint operator $M$ on $\mathcal{H}$%
, and let $L_{1}$ and $L_{2}$ be bounded self-adjoint operators on $%
\mathcal{H}$ and $\mathcal{K}$, respectively, such that  
$[U,L_{1}\otimes \boldsymbol{1}+\boldsymbol{1}\otimes L_{2}]=0$. Assume that $%
{\mathcal{M}}$ is repeatable or satisfies the Yanase condition.
Then $\left[L_{1},M\right] =0$.}

%begin{proof}
\noindent{\em Proof.}
We choose orthonormal bases $\{\varphi _{\mu \rho }\}$ and $\left\{ \phi _{\mu \sigma }\right\} $
of eigenstates of $M$ and $Z$, respectively (with $\rho $,$\sigma$ as degeneracy parameters). 
The most general unitary coupling $U$ that constitutes a measurement of $M$
then takes the form
\begin{equation}
\varphi _{\mu \rho }\otimes \phi \overset{U}\longrightarrow \sum\limits_{\sigma
}\varphi _{\mu \rho \sigma }^{\prime }\otimes \phi _{\mu \sigma },
\end{equation}%
where 
$\{\varphi _{\mu \rho \sigma }^{\prime }\}$ in $\mathcal{H}$ is an arbitrary set of states such that 
$\sum_{\sigma}\left\Vert \varphi _{\mu \rho \sigma }^{\prime }\right\Vert^{2}=1$.   Implementing the conservation law
(given by $\left[ U,L\right] =0)$ we may now write the matrix elements of $L$ in the following way:%
\begin{equation}
\left\langle \varphi _{\mu ^{\prime }\rho ^{\prime }}\otimes \phi |L\varphi
_{\mu \rho }\otimes \phi \right\rangle =\sum\limits_{\sigma ,\sigma ^{\prime
}}\left\langle \varphi _{\mu ^{\prime }\rho ^{\prime }\sigma ^{\prime
}}^{\prime }\otimes \phi _{\mu ^{\prime }\sigma ^{\prime }}|L\varphi _{\mu
\rho \sigma }^{\prime }\otimes \phi _{_{\mu \sigma }}\right\rangle .
\label{WAY_matrix}
\end{equation}%
The additivity of $L$ and the assumption that $\phi $ is normalized entails
that (\ref{WAY_matrix}) can be written%
\begin{align}\label{WAY_sum}
&\left\langle \varphi _{\mu ^{\prime }\rho ^{\prime }}|L_{1}\varphi _{\mu\rho }\right\rangle 
+\left\langle \varphi _{\mu ^{\prime }\rho ^{\prime}}|\varphi _{\mu \rho }\right\rangle \left\langle \phi |L_{2}\phi
\right\rangle\nonumber \\
&\hspace{1cm}=\sum\limits_{\sigma ,\sigma ^{\prime }}\left[ \left\langle \varphi _{\mu
^{\prime }\rho ^{\prime }\sigma ^{\prime }}^{\prime }|L_{1}\varphi _{\mu
\rho \sigma }^{\prime }\right\rangle \left\langle \phi _{\mu ^{\prime
}\sigma ^{\prime }}^{\prime }|\phi _{\mu \sigma }^{\prime }\right\rangle
+\right.\\ 
&\hspace{2.5cm} \left.\left\langle \varphi _{\mu ^{\prime }\rho ^{\prime }\sigma ^{\prime
}}^{\prime }|\varphi _{\mu \rho \sigma }^{\prime }\right\rangle \left\langle
\phi _{\mu ^{\prime }\sigma ^{\prime }}|L_{2}\phi _{\mu \sigma}\right\rangle \right] . \nonumber 
\end{align}%
By the orthogonality of pointer eigenstates, $\left\langle \phi _{\mu
^{\prime }\sigma ^{\prime }}^{\prime }|\phi _{\mu \sigma }^{\prime
}\right\rangle =0$ for $\mu \neq \mu ^{\prime }$; examination of each of the
remaining terms in the sum in the above expression tells us that these
vanish if one of the following conditions holds:\\
(a) $\left\langle \varphi _{\mu ^{\prime }\rho ^{\prime
}\sigma ^{\prime }}^{\prime }|\varphi _{\mu \rho \sigma }^{\prime
}\right\rangle =0$ for $\mu \neq \mu ^{\prime }$;\\
(b) $\left\langle \phi _{\mu
^{\prime }\sigma ^{\prime }}|L_{2}\phi _{\mu \sigma }\right\rangle =0$ for $\mu \neq \mu ^{\prime }$.\\ 
Condition (a) is satisfied whenever the measurement is repeatable.
Condition (b) is satisfied exactly when the eigenspaces of the pointer observable are invariant 
under the action of $L_{2}$, i.e when $[L_{2},Z]=0$, so that the
measurement satisfies the Yanase condition. If either of these are satisfied, 
then the right hand side of (\ref{WAY_sum}) is zero, 
and thus the left hand side must vanish also. 
Clearly the second term on the left hand side vanishes due to the orthogonality of the eigenstates
of $M$, and the first vanishes if and only if $L_{1}$ leaves $M$--eigenspaces invariant, i.e. if and only if 
$\left[ L_{1},M\right] =0$.
%\end{proof}

We interpret the theorem as follows: if ${\mathcal{M}}$ is a measurement of $M$ and $\left[
L_{1},M\right] \neq 0$, then the conservation of $L$ entails
that ${\mathcal{M}}$ must violate both repeatability and the Yanase
condition, in accordance with the expectation that emerged in
the previous section. 

As the proof shows, the commutativity of $M$ with $L_1$ follows from the condition (a), which is in fact
a weakening of the repeatability requirement as it merely requires the distinguishability of the 
post-measurement states of the system. Repeatability is obtained by assuming that the $\varphi_{\mu\rho\sigma}'$
are eigenvectors of $M$.  In  \cite{miyadera}  it has been shown that the distinguishability of
the post measurement object states on one hand and of the 
post measurement apparatus states on the other are
subject to a WAY-type trade-off relation. There the distinguishability is quantified by a measure of \emph{fidelity},
and the measurement inaccuracy is manifested by final pointer states having non-maximal fidelity.

We note that a result of the form of the above theorem (i.e. using the weakened form of repeatability or the Yanase 
condition to derive the commutativity of the observable to be measured with the conserved quantity) has been 
proved by Beltrametti et al in 1990 \cite{BCL} for the special case of {\em minimal unitary measurements}, for which the
spectra of both the measured observable and pointer are nondegenerate.

As noted above, the violation of the Yanase condition can be understood 
as disallowing accurate and repeatable measurements of the apparatus observable (since 
this observable is now subject to the same limitations as prescribed by the WAY theorem). 
We also observed that the repeatability of pointer measurements is required for ensuring
stable and reproducible measurement records. Hence, even if repeatability is sacrificed at the object level,
it would seem indispensable at the level of the pointer measurement, thereby enforcing fulfillment
of the Yanase condition. Thus we argue that no ``measurement''
violating the Yanase condition may be called a measurement at all. One may talk only 
of information transfer between system and apparatus and must also consider
how this information can be finally extracted. This conclusion applies to the class 
of pointer observables that are subject to the WAY theorem.

\subsection{Technical developments}\label{subsec:tech}

As demonstrated in a footnote in  \cite{way},
the case of $L_2$ being unbounded can be
incorporated into the proof in a natural way. This is achieved by using the unitary
operators $V(t)=\exp{(itL)}$ and $V_{i}(t)=\exp{(itL_i)}$ (with $i=1,2$, $t\in\mathbb{R}$) 
and noting that $V(t)=V_{1}(t)\otimes V_{2}(t)$. Then one can
follow the previous line of proof, replacing the original operators with their exponentiated 
forms, and exploiting the boundedness of $L_1$. 

Ghirardi et al \cite{Ghi-etal} have extended the WAY theorem to the case where $L_1$ may be 
unbounded, but all eigenvectors of $M$ are contained in the domain of $L_1$. The measurement
is still stipulated to be repeatable. They note that their theorem constrains the feasibility of 
repeatable measurements of a component of the orbital angular momentum observable in the presence
of the conservation of another angular momentum component for the system plus apparatus. 
Yet their extension still does not 
cover some physically important cases, namely, those involving observables with continuous 
spectra.

\section{WAY-type Limitation for Approximate Measurements}\label{sec:approx}

Wigner's paper \cite{wigner} not only demonstrated the strict impossibility of
accurate and repeatable measurements given the conservation law,
but also delineated a means by which approximate measurements with approximate
repeatability could be recovered. It is also the case, as demonstrated by Araki and Yanase,
that this positive part of Wigner's example can be extended to a 
much more general class of observables and conserved quantities. Here we 
describe further developments in this area, examine WAY-type limitations for approximate 
measurements, and discuss how approximate repeatability also follows a trade-off relation 
with the size of the apparatus in certain circumstances. This helps to elucidate further the crucial role 
of the Yanase condition in discussions of WAY-type limitations to quantum measurements.

In the case where $\left[L_{1},M\right] \neq 0$, the limitation given by the WAY theorem
can thus be re-expressed more quantitatively: 
There are approximate measurements of $M$,  with some degree of approximate repeatability, 
which satisfy the Yanase condition, but where good approximations are achieved at the price of
requiring a large apparatus, quantified by the magnitude of $\langle \phi
|(L_{2})^{2}\phi \rangle $.

\subsection{Overview of results}\label{subsec:overview}

Yanase \cite{yanase-opt} derives an ``optimal'' lower bound for the probability of the measurement
failing to be accurate and repeatable; he considers  measurements of a spin component $S_{x}$ where the
conserved quantity is $S_{z}+J_{z}$, with 
$J_{z}$ the $z$-component of the apparatus' (unbounded) angular momentum . The pointer observable is chosen so that it
commutes with $J_{z}$. 
In this case, the lower bound for the probability of the apparatus  malfunctioning
is given by $[8\langle \phi | (J_{z})^{2}\phi \rangle]^{-1}$. This bound was also illustrated
by Ghirardi \cite{Ghirardi} for rotationally invariant Hamiltonians. Yanase's result, though
claimed to be ``optimal'', still only considered terms up to second moments in $(J_{z})$, and thus
optimality was not proven rigorously.  This was pointed out by Ozawa \cite{ozawacon}
who obtained a sharper, tight bound without approximations.

Ghirardi et al \cite{Ghi-etal} have considered the case where measurement errors arise from the non-orthogonality
of the final apparatus states. They consider both ``distorting'' and ``non-distorting'' (yet still repeatable) measurements. 
They derive lower bounds on the probability of the ``malfunctioning'' of the apparatus, and even consider
the role that large apparatus size has in reducing these probabilities. However, they do not establish the
{\em necessity}  of a large apparatus for good measurements; they merely assume that the error probabilities
can be made small by increasing the expectation of the square of the apparatus part of the conserved quantity.

\subsection{Ozawa's trade-off inequality}\label{subsec:Ozawa}

Ozawa \cite{ozawacon} develops an alternative formulation of the WAY theorem. He introduces a measure of noise to quantify measurement inaccuracy, and shows that this
has a lower bound that can be decreased provided the variance of the apparatus'
conserved quantity is increased. This trade-off inequality follows as an application of the 
Cauchy--Schwarz inequality.

Given a measurement $\mathcal M$ that is to serve as an approximate determination of 
an observable $M$, the {\em noise operator} is defined as the difference 
$N:=Z(\tau )-M$, where $Z(\tau )$ represents the Heisenberg-evolved pointer 
observable after the interaction period $\tau $. A measure of  \textit{noise} is then given as
$\epsilon (\varphi )^{2}:=\left\langle
\varphi \otimes \phi |N^{2}\varphi \otimes \phi \right\rangle \equiv \langle
N^{2}\rangle $. Clearly $\epsilon (\varphi )^{2}\geq (\Delta N)^{2}$. A global
measure of \emph{error} can be provided 
by taking the supremum over all (normalized) input states $\varphi$ of the 
quantity $\epsilon (\varphi )^{2}$, i.e. $\epsilon^{2}:=\sup_\varphi\epsilon(\varphi)^{2}$. 
This quantity should be finite for any measurement that would qualify as an approximate
determination of $M$.
Then the uncertainty relation
entails
\begin{equation}
\epsilon^{2}\geq
\epsilon (\varphi )^{2}\geq \frac{1}{4}\frac{\left\vert \left\langle \left[
Z(\tau )-M,L_{1}+L_{2}\right] \right\rangle \right\vert ^{2}}{(\Delta L)^{2}}%
,  \label{Ozawa_error}
\end{equation}
where it is found that $(\Delta L)^{2}= (\Delta _{\psi }L_{1})^{2}+(\Delta _{\phi }L_{2})^{2}$.
The measurement is accurate if and only if $\epsilon=0$.

Thus, if the Yanase condition $(\left[ Z,L_{2}\right] =0)$ is satisfied and the interaction
obeys the conservation law, then all that remains in the numerator is 
$\left\vert \left\langle \left[ M,L_{1}\right] \right\rangle \right\vert ^{2}$. If this is
zero then there is no lower bound on the measurement accuracy, in accordance
with the findings of WAY. 

In the case that $\left\vert \left\langle \left[
M,L_{1}\right] \right\rangle \right\vert ^{2}$ is non-zero but finite, then
clearly if $(\Delta L)^{2}$ becomes large the lower bound on the inaccuracy decreases.
Furthermore, since the initial system state is arbitrary, only by fixing $\phi$ 
such that $(\Delta _{\phi }L_{2})^{2}$ is large may one increase the accuracy of the measurement, thus
establishing the necessity of a large apparatus variance for good measurements.

It is also worthwhile investigating the case of a measurement scheme $\mathcal{M}$ that satisfies 
neither the Yanase condition nor the commutativity condition $[M,L_{1}]=0 $ but is such that 
the bound on the right hand side of (\ref{Ozawa_error}) vanishes; thus,
$\left[ Z(\tau ),L_{1}+L_{2}\right] =\left[ M,L_{1}\right]=U^*\left[ Z,L_{2}\right] U $, by
the conservation law. 
This is clearly satisfied if $\mathcal M$ happens to be accurate, $\epsilon=0$.
Such a measurement scheme allows for perfectly accurate transfer of information
from system to apparatus, and demonstrates the necessary failure of the 
Yanase condition for this to be achieved. 

\subsection{Trade-off relation for repeatability}\label{subsec:repeatability}

Ozawa \cite{repeatable}  has proved that observables with a continuous spectrum do not admit 
any repeatable measurements. This holds regardless of whether there are additive conserved 
quantities or not. In order to describe repeatability properties of measurements of such observables, 
it is therefore necessary to have notions of approximate repeatability, and methods for quantifying 
how repeatable a measurement is. One approach to  weaken condition \eqref{rep} \cite{Davies,OQP}. We will explain and use this  in Sec. \ref{subsec:Stein/Shimony} in the context
of a measurement model.

Here we introduce a different intuitive quantification of repeatability that is somewhat similar to
the construction of the noise operator. With this we can generically describe how repeatable a 
measurement is by utilizing a commutation relation with the conserved quantity. We define:
\begin{equation}
\ \mu (\varphi )^{2}:=\langle \varphi \otimes \phi |(M(\tau )-Z(\tau
))^{2}\varphi \otimes \phi \rangle \text{;}  \label{rep_meas}
\end{equation}%
intuitively if this expectation is small, then the difference between the
measured observable and the time-evolved system observable is small, and
hence the measurement should display some level of repeatability.\ A state
independent measure of repeatability may thus be defined as $\mu ^{2}:=\sup
\mu (\varphi )^{2}$, yielding 
\begin{equation}
\mu ^{2}\geq\sup_{\varphi } \frac{1}{4}\frac{\left\vert \left\langle \left[ M(\tau )-Z(\tau
),L_{1}+L_{2}\right] \right\rangle \right\vert ^{2}}{(\Delta_{\varphi }
L_{1})^{2}+(\Delta _{\phi }L_{2})^{2}} . \label{app_rep}
\end{equation}%
If the Yanase condition is satisfied, then $\left[ Z(\tau),L_{1}+L_{2}\right] =0$ and so
\begin{equation}
\mu ^{2}\geq\sup_{\varphi } \frac{1}{4}\frac{\left\vert \left\langle \left[ M(\tau ),L_{1}+L_{2}\right] \right\rangle \right\vert ^{2}}{(\Delta_{\varphi }
L_{1})^{2}+(\Delta _{\phi }L_{2})^{2}} ,
\end{equation}%
which demonstrates that good repeatability may also be achieved when 
$(\Delta _{\phi }L_{2})^{2}$\ is large. This condition becomes a necessity when
$\left[ M,L_{1}\right] $ is non-zero.

\section{``WAYs Out"}\label{sec:wayout}

If an observable we wish to measure does not commute with an additive conserved quantity, 
we have seen that one may still obtain perfectly accurate information transfer between system 
and apparatus despite the WAY theorem. Here we note some realizations in which this is achieved, 
and show explicitly that these models violate both repeatability and the Yanase condition.

\subsection{Ohira and Pearle}\label{Ohira and Pearle}

Ohira and Pearle \cite{ohira/pearle} provide a ``WAY-out''
of the limitation arising from the WAY theorem via a model in which both the object 
and the probe are given as spin-$\frac{1}{2}$ systems. The measurement
coupling is generated by a rotationally invariant Hamiltonian of the form 
$H=(\mathbf{S}+\mathbf{J})\cdot (\mathbf{S}+\mathbf{J})$.

We proceed under the notation that $\psi _{\pm }$ represent both $S_{z}$ and  $J_{z}$ eigenstates,
and $\phi =\psi_+$. The evolution takes the form (with the interaction period $\tau = \pi /2$):
\begin{equation}
\label{O+H_2}
\begin{aligned}
(\psi _{+}+\psi _{-})\otimes \phi &\longrightarrow &(-\psi _{+})\otimes
(\psi _{+}+\psi _{-}), \\
(\psi _{+}-\psi _{-})\otimes \phi &\longrightarrow &(-\psi _{+})\otimes
(\psi _{+}-\psi _{-}) .
\end{aligned}
\end{equation}
Here  the appropriate pointer observable is $Z=J_{x}$. This model is not repeatable, and also
violates the Yanase condition.
 
Recalling equations \eqref{mmt1} and \eqref{mmt2} which appeared on Wigner's final page,
we see that these have precisely the same form as (\ref{O+H_2}), apart from an inconsequential difference of initial pointer states.

Our analysis of this model of Ohira and Pearle coincides with that of Wigner's last page (Sec. \ref{subsec:alternative}). 
They point out that this model has demonstrated that if repeatability is not insisted upon, one may achieve an accurate 
measurement despite the restrictions of the WAY theorem.  However, as we have seen, the theorem does not stipulate 
\emph{any} limitation to the accuracy (of information transfer) when both the repeatability and Yanase conditions are
 violated, as is the case here. This is precisely the setting in
which perfect accuracy is achievable, and this model of Ohira and
Pearle is therefore fully in accordance with the WAY theorem as we have given it.

Ohira and Pearle's aim was to expose and correct a common misreading of the WAY theorem as prohibiting accurate measurements
in the presence of an additive conserved quantity. This prohibition, they show, is removed at the expense of giving up the repeatability 
of the measurement. We know now that in addition the Yanase condition has to be violated as well.

 Ozawa's inequality \eqref{Ozawa_error} shows how the zero-error measurement can be achieved; the condition 
 for vanishing lower bound for the error takes the form 
$U^*\left[ Z,L_{2}\right] U=\left[ M,L_{1}\right]$. In this model,  it is easily verified that 
$U^{\ast }\boldsymbol{1}\otimes S_{y}U=S_{y}\otimes \boldsymbol{1}$, which 
indeed entails that the expectation value in 
the numerator of Ozawa's inequality vanishes. 

\subsection{The SWAP Map Example}\label{sebsec:SWAP}

Following the work of Wigner and Ohira and Pearle, we note that these ``WAYs out''  are both examples
of a remarkably simple structure. They violate both repeatability and the Yanase condition, and whenever the initial system
state is an eigenstate of the observable to be measured, both take the form of an unentangled (product) state after the 
unitary interaction. It is known \cite{busch_unitary} that the only non-entangling unitary operators $U$ on 
$\mathcal{H}_1 \otimes \mathcal{H}_2 $ are either of the form: (i) $U(\varphi \otimes \phi) = (V \varphi) \otimes (W \phi)$ 
(with $V$ and $W$ unitary on $\mathcal{H}_1$ and $\mathcal{H}_2$ respectively), or 
(ii) $U(\varphi \otimes \phi) = (V_{21} \phi) \otimes (W_{12} \varphi)$ with $V_{21}: \mathcal{H}_2 \rightarrow \mathcal{H}_1$ 
and $W_{12}: \mathcal{H}_1 \rightarrow \mathcal{H}_2$ surjective 
isometries. This latter scenario is only possible if $\dim{\mathcal{H}_1}=\dim{\mathcal{H}_2}$ (with the dimension possibly infinite).

One of the simplest examples of a non-entangling unitary map (which is of type (ii), see above)
is provided by the \emph{SWAP} map $U_{S}$ on $\mathcal{H}\otimes\mathcal{H}$, defined by
$U_{S}(\varphi\otimes\phi)=\phi\otimes\varphi$.
If this unitary map is to be used in the context of a measurement, we see that \eqref{PRC} takes the form
 $ \left\langle \varphi |\Esf(X)\varphi \right\rangle = \left\langle \varphi |\Esf^{Z}(f^{-1}(X))\varphi \right\rangle$ 
 (for all $\varphi \in{\mathcal{H}}$), which can be satisfied if $\Esf=\Esf^{M}=\Esf^{Z}$, and hence $Z=M$. 
 This also respects any conservation law that is additive and where each non-trivial operator in the sum 
 takes the same form. The noise operator is given as $N=U^{\ast }(\boldsymbol{1}\otimes Z)U-M\otimes \boldsymbol{1=}Z\otimes 
\boldsymbol{1}-M\otimes \boldsymbol{1}$. Thus, since we have chosen $Z=M$, the noise operator $N$ 
vanishes and we have a perfectly accurate  information transfer between system and apparatus. 
However, as the SWAP map violates
the Yanase condition, there remains the problem of recovering this information from the pointer observable.

\section{Position Measurements Obeying Momentum Conservation}\label{sec:pos}

Many of the observables that make up a coherent and complete view of (quantum) physical
reality are not of the class that have been discussed thus far. Technical difficulties arise in the context of 
unbounded operators with continuous spectrum, position and (linear) momentum being two noteworthy 
examples. However if one wishes
for a comprehensive understanding of WAY-type limitations to the measurability of physical quantities, 
it is critical to understand the fundamental case of position measurements that obey momentum 
conservation. In this
section we discuss some results that have been obtained in this context. Any WAY-type theorem for 
these observables will have to take into account Ozawa's result that as a continuous quantity,
position cannot be measured repeatably.

In \cite{Loveridge/Busch} the present authors have provided strong evidence for
the existence for such a theorem in the position--momentum case. They
demonstrate that a model put forward by Ozawa claiming to demonstrate no
WAY-type restriction is flawed. The model of Ozawa
satisfies the Yanase condition, and one can show that only in the limit of
the pointer preparation becoming a delta-function may the inaccuracy tend to
zero, which comes at the expense of the apparatus' momentum distribution having a
large width (suitably defined). Furthermore \cite{Loveridge/Busch} provides a model that
explicitly violates the Yanase condition, where arbitrarily accurate and repeatable
measurements may still be achieved without resorting to a size constraint on the apparatus.

\subsection{A General Argument}\label{gen}

It is again possible to implement the Ozawa inequality (\ref{Ozawa_error})
to obtain a general argument in favour of WAY-limitations in the continuous
unbounded case when the Yanase condition is satisfied. 
The form of the position--momentum commutator allows 
the supremum on the right-hand side of (\ref{Ozawa_error}) to be taken in the following way:
\begin{equation}
\epsilon ^{2}\geq \frac{1}{4}\frac{1}{\inf_{\varphi }(\Delta _{\varphi
}P)^{2}+(\Delta _{\phi }P_{\mathcal{A}})^{2}}=\frac{1}{4(\Delta _{\phi }P_{
\mathcal{A}})^{2}}.
\end{equation}%
with $(\Delta _{\varphi}P)^{2}$ and $(\Delta _{\phi }P_{\mathcal{A}})^{2}$ the
variance of the momentum in the system and apparatus respectively.
This bound allows for an increase in accuracy only  when $(\Delta _{\phi }P_{%
\mathcal{A}})^{2}$ is large, establishing the necessity of large apparatus size for
good measurements. 

Precisely the same bound arises when one considers the
repeatability (defined in (\ref{app_rep}));%
\begin{equation}
\mu ^{2}\geq \frac{1}{4(\Delta _{\phi }P_{\mathcal{A}})^{2}}.
\end{equation}%
This provides an indication that good repeatability can indeed be achieved if (and only if) there is a large momentum variance in the probe.

Notice that the non-zero lower bounds to both accuracy and repeatability arise after explicit implementation of the Yanase
condition, $[Z,P_{\mathcal{A}}]=0$. If we relinquish this condition, there is nothing that would prevent $[
Z(\tau )-Q,P+P_{\mathcal{A}}] $ from vanishing. Indeed this would be the case in any model
where one could choose the pointer observable as the apparatus' position, $Q_\mathcal{A}$.

In the position--momentum case, the role of the Yanase condition must
be considered very carefully. Previously (in the case where the WAY theorem certainly applied)
we argued for the Yanase condition by applying the WAY theorem to the measurement of the 
pointer, of which we demanded accurate and repeatable measurements.
However, since no such theorem has been proven in the continuous/unbounded case, one must 
be more tentative when stipulating this condition, and it may be considered as a precautionary manoeuvre. The
models discussed in  \cite{Loveridge/Busch}, as well as the above model-independent relations point
in the direction of a WAY-type limitation if the Yanase condition is satisfied and no such obstruction if it is not.

The last conclusion (of ``no obstruction") contrasts, perhaps somewhat surprisingly, the WAY theorem for accurate 
measurements: Within the realm of that theorem, it is not sufficient to violate the Yanase condition in order to lift the
obstruction against perfect accuracy and repeatability.
The fact that no size constraint is required for good measurements of position if the pointer observable is a position
itself can be understood by considering the lower bounds in equations (\ref{Ozawa_error}) and (\ref{app_rep}): 
If the object position does not change during the interaction, $M(t)=M=Q$, and the pointer is $Z=Q_{\mathcal A}$,
the lower bounds become zero in both cases since the commutator of the noise operator $N=Q_{\mathcal A}(t)-Q$ 
with the conserved quantity $L_1+L_2=P+P_{\mathcal A}$ vanishes identically. This is a consequence of the fact that
$[Q_{\mathcal A}(t),P+P_{\mathcal A}]=i \boldsymbol{1}=[Q,P]$. Such cancellation of commutators living on different
Hilbert spaces can only arise for pairs of observables with constant commutators. 

It is not known whether, under 
violation of the Yanase condition, there exist measurements of position that are fully accurate, and repeatable 
to a good approximation. It is also an open problem whether, again with giving up the Yanase condition, 
approximate spin measurements obeying angular momentum conservation are possible with good repeatability
properties, without any constraint on the size of the apparatus.

\subsection{The Problem of Stein and Shimony}\label{subsec:Stein/Shimony}

In 1979 Stein and Shimony \cite{Stein-Shimony} posed a problem concerning the possibility of realizing a 
two-valued (and hence coarse-grained) position measurement that respects the conservation of momentum. 

This problem takes the form of whether there
exists a non-zero function $\phi \in{L^2(\mathbb{R})}$ and unitary operator 
$U:L^2(\mathbb{R}^2) \rightarrow L^2(\mathbb{R}^2)$ that commutes with the shift operators (defined by 
$T_{t}(g)(x,y)=g(x+t,y+t)$ for $g \in{ L^2(\mathbb{R}^2)}$ and $t \in{\mathbb{R}}$)) and satisfy:
\begin{align*}
&\supp{[U(\varphi \otimes \phi)]} \subseteq\mathbb{R}^{+}\times \mathbb{R}^{+}\ {\rm if}\ \supp{\varphi} \subseteq\mathbb{R}^{+},\\
&\supp{[U(\varphi \otimes \phi)]}\subseteq\mathbb{R}^{-}\times \mathbb{R}^{-}\ {\rm if}\ \supp{\varphi} \subseteq\mathbb{R}^{-},
\end{align*}
where $\varphi\in L^2(\mathbb R)$. With the pointer being a two-valued, discretized position observable,
this coupling necessarily violates the Yanase condition. The condition that the unitary $U$ commutes with 
$T_{t}$  is a mathematical expression of the conservation of the total momentum $P + P_{\mathcal{A}}$.

Here we provide a position measurement scheme \cite{OQP} that approximately satisfies the above 
requirements with the quality of the approximation becoming arbitrarily good as the value of the 
coupling parameter $\lambda$ becomes large. The momentum--conserving unitary operator $U$ which describes the 
interaction is given by
\begin{equation}
U=\exp \left[ -i\frac{\lambda }{2}\bigl((Q-Q_{\mathcal{A}})P_{\mathcal{A}
}+P_{\mathcal{A}}(Q-Q_{\mathcal{A}})\bigr)\right] ,
\end{equation}
where for example we have written $(Q-Q_{\mathcal{A}})P_{\mathcal{A}
}$ as shorthand for $(Q\otimes \mathbf{1}-\mathbf{1} \otimes Q_{\mathcal{A}})\mathbf{1} \otimes P_{\mathcal{A}
}$. The pointer observable is given as $Q_{\mathcal{A}}$,
and the measured observable $\Esf$ [Eq.~(\ref{PRC})]
is of the form $\Esf(X)=\chi _{X}\star e(Q)$, if the scaling
function $f$ is chosen such that $f^{-1}(X)=(1-e^{-\lambda })X$. Here $\chi_X$ represents the characteristic set function. The
probability density $e=e^{(\lambda )}$  
depends on $\lambda $ in the following way: 
\begin{equation}
e^{(\lambda )}(q)=(e^{\lambda }-1)\left\vert \phi (-q(e^{\lambda}-1))\right\vert ^{2}. 
 \end{equation}

In order to answer the question of Stein and Shimony, we first recast the conditions that need to 
be satisfied as follows. Firstly, the
measurement must satisfy a stronger form of the probability reproducibility condition called the
\emph{calibration condition}, which requires that if the initial state is localized in the positive
(or negative) half line, then this result is shown on the pointer with certainty. We shall denote
the spectral measures of $Q$ and $Q_{\mathcal{A}}$ by $\mathsf{Q}$ and $\mathsf{Q}_{\mathcal{A}}$ 
respectively. Allowing for some error, this may be written (for $\alpha >0$)

\begin{equation}\label{calib}
\left\langle \Psi _{\tau} |\mathbf{1}\otimes \mathsf{Q}_{\mathcal{A}}[-\alpha,\infty)
\Psi _{\tau}\right\rangle =1
\end{equation}
if $ \supp{\varphi} \subseteq \lbrack 0,\infty )$, and we show that $[
-\alpha ,\infty )$ can become arbitrarily close to $[0,\infty )$ if $\lambda $
is made suitably large.

The second requirement is that of repeatability, which we give as a slightly modified version
of \eqref{rep} whereby the immediate subsequent measurement is of the observable $Q$.  
This takes the form (with $\beta>0$)
\begin{multline}
\left\langle \Psi _{\tau }|\mathsf{Q}[-\beta,\infty) \otimes 
\mathsf{Q}_{\mathcal{A}}\mathbb{(R}^{+}\mathbb{)}\Psi _{\tau }\right\rangle
=\\ \left\langle \Psi _{\tau }|\boldsymbol{1}\otimes \mathsf{Q}_{\mathcal{A}}%
\mathbb{(R}^{+}\mathbb{)}\Psi _{\tau }\right\rangle = \left\langle \varphi |\mathsf{E}(%
\mathbb{R}^{+})\varphi \right\rangle, \label{SSrep}
\end{multline}
where the last equality results from the probability reproducibility
condition. We shall show that this may be satisfied for all $\varphi$ and that $\beta$ can be 
made arbitrarily small.

We shall make the immediate specification that the initial state wave function 
$\phi $ of the apparatus be supported on a fixed finite interval of width $2{\ell}$ around the
origin; $ {\supp\phi} =[-{\ell},{\ell}]$. Therefore the distribution $
e^{(\lambda )}$ is supported on the $\lambda $-scaled interval $[-\delta
,\delta ]$, with $\delta =\ell / (e^{\lambda }-1)$.

After some manipulation the calibration requirement \eqref{calib} takes the form
\begin{equation}
\int_{0}^{\infty} \left\vert \varphi (q)\right\vert ^{2}\chi _{\lbrack -\alpha ^{\prime
},\infty )}\ast e^{(\lambda )}(q)dq=1
\end{equation}
with $\alpha ^{\prime }=f(\alpha)$.
Thus we require $\chi _{\lbrack -\alpha ^{\prime },\infty
)}\ast e^{(\lambda )}(q)=1$ for all $q \geq{0}$ and so
\begin{equation}
\int_{-\infty }^{\alpha ^{\prime }+q}e^{(\lambda )}(y)dy=1,
\end{equation}
which is satisfied if $q\geq\delta -\alpha ^{\prime }$. The smallest $\alpha ^{\prime }$ consistent with 
this constraint occurs when $\alpha ^{\prime
}=\delta $, and so 
$\alpha ={\ell}e^{-\lambda}$. Therefore we see that indeed $\alpha
\rightarrow 0$ as $\lambda \rightarrow \infty $. It must also be
shown that the same behaviour emerges in the case when $\supp{\varphi \subseteq{(-\infty,0]}}$ 
but we omit this essentially identical calculation, and this completes the proof.

We now address the repeatability requirement. 
Writing \eqref{SSrep} in integral form and rearranging, we see that
\begin{equation}
\int \left\vert \varphi (q)\right\vert ^{2}(\chi _{\lbrack \mathbb{-}\beta
,\infty )}(q)-1)\chi _{[0, \infty)}\ast e^{(\lambda)}(q)dq=0,
\end{equation}
and so
\begin{equation}
\chi _{( \mathbb{-}\infty ,-\beta )}(q)\int_{-\infty
}^{q}e^{(\lambda )}(y)dy=0.
\end{equation}
This expression certainly vanishes if $q\ge -\beta $. When $q< -\beta $, recalling that 
$\supp{e^{(\lambda )}}=[-\delta ,\delta ]$, we see that if $-\delta \geq -\beta $
(and thus $\beta \geq \delta $) then the integral vanishes.  Since we are
looking for the smallest $\beta $ for which this may be satisfied, we
choose $\beta =\delta =\ell/(e^{\lambda} -1)$. Therefore in the large 
$\lambda $ limit, $\beta $ is arbitrarily small, showing that\ arbitrarily
good repeatability may be achieved. Due to the symmetry of the support of $e^{(\lambda )}$, 
it follows that arbitrarily
good repeatability holds also for the $\mathbb{R}^-$ outcome on the pointer. 

Although this model provides only an approximate solution to the problem of Stein and Shimony,
we note that from an operational perspective this does not differ from an exact solution. Since the 
accuracy and approximate repeatability can be made arbitrarily good by simply tuning the coupling 
parameter, in any experimental realization this could not be distinguished from a measurement in which 
perfect accuracy and repeatability can be achieved. This does not require a large momentum spread in 
the probe, and it has been shown that the present model indeed presents an  approximate measurement 
scheme for the 
full position observable $Q$, with arbitrarily good accuracy and repeatability properties \cite{Loveridge/Busch}.

\section{Concluding Remarks}\label{sec:conclusion}

The WAY theorem, with its generalizations, is applicable to a large class of physically important scenarios. 
In any situation in which, for example, spin or angular momentum is the relevant observable, the measurement 
accuracy is likely to be hampered by a WAY-type constraint. When considering the manipulation of 
individual quantum objects using other small objects as `apparatus', it may not be possible to fulfill the 
requirement of large variance of the apparatus part of the conserved quantity. Such scenarios do occur in  
quantum information processing and quantum control. 
Ozawa and coworkers \cite{conquant} have in fact demonstrated a limitation to the realizability of quantum 
logic gates insofar as the observables involved are subject to the WAY theorem. This has led to an increased 
awareness that attention has to be paid to the presence of conserved quantities in the design of quantum gates.

In the case of position measurements that obey momentum conservation, no WAY-type 
obstruction exists if one asks only for a measurement of the relative distance between the object and a 
``reference system''. In this case, when the reference system is provided by part of 
the apparatus,  the measured observable can be given as the 
relative position. As is alluded to in \cite{AR}, it appears that there is a link to the theory of superselection 
rules and quantum reference frames (see, e.g. \cite{qfr}), 
which has been the subject of much interest and investigation recently. This possible link opens up an 
avenue that requires further systematic study.
\\

\noindent
\textbf{Acknowledgments.} Thanks are due to Rebecca Ronke and Tom Potts for
many helpful discussions and careful reading of drafts of this manuscript.
This work was supported by EPSRC UK.

\section*{Appendix: Reconstructing Wigner's last page}\label{sec:last-page}

In this appendix we shall carefully reconstruct the argument that appears 
on the final page of Wigner's 1952 paper \cite{wigner}. Although Wigner's work
is succinct and simple, the lack of detailed calculations makes reproducing his conclusions somewhat
harder work than one might imagine. We also present some subtly different arguments from
those found in the original work.

Wigner restricts his consideration to the case where the post-interaction states are
of product form (unentangled) in the system--apparatus Hilbert space, and he makes the choice that
 the initial apparatus state $\phi$ be an eigenstate 
of $S_z$ with eigenvalue zero. He writes
\begin{eqnarray}
(\psi _{0}+\psi _{1})\otimes \phi &\longrightarrow
&\sum\limits_{i=0}^{1}\psi _{i}^{\prime }\otimes \sum \phi _{j}^{\prime },
\label{wig1} \\
(\psi _{0}-\psi _{1})\otimes \phi &\longrightarrow
&\sum\limits_{i=0}^{1}\psi _{i}^{\prime \prime }\otimes \sum \phi
_{j}^{\prime \prime },  \label{wig2}
\end{eqnarray}%
with $\psi _{i}^{\prime }$ and $\psi _{i}^{^{\prime \prime }}$
representing un-normalized $S_{z}$ eigenstates. In order that Wigner's 
analysis be compelling, we must assume $\phi _{j}^{\prime }$
and $\phi _{j}^{^{\prime \prime }}$ to be eigenstates of the apparatus' angular
momentum, $J_{z}$. The reason for this choice will become clear shortly; 
this is the only way in which consistency with the conservation law can be maintained. 
We omit summation indices on the apparatus Hilbert space since it is assumed to run
to infinity. However, the number of non-zero terms in this
expansion is dramatically reduced due to the choice of initial apparatus state and the conservation law;
the left hand side of (\ref{wig1}) contains a superposition of $S_{z}$ eigenstates, and thus a 
superposition of states containing zero and one ``unit'' of the
conserved quantity. The right hand side cannot, then, contain more than
one such unit. 

In order to correspond to Wigner's analysis, we proceed under the restriction that {0} be the 
lowest eigenvalue for the apparatus' conserved quantity, and from here it follows that (
\ref{wig1}) and (\ref{wig2}) take on a much simpler forms. With $\phi=\phi _{0}$ and dropping all terms with
the apparatus containing two or more units of the conserved quantity, we have
\begin{multline}
(\psi _{0}+\psi _{1}) \otimes  \phi _{0}  \longrightarrow 
 \psi _{0}^{\prime}\otimes \phi _{0}^{\prime }+\psi _{0}^{\prime }\otimes
\phi _{1}^{\prime }+\psi _{1}^{\prime }\otimes \phi _{0}^{\prime }+\psi
_{1}^{\prime }\otimes \phi _{1}^{\prime } ,    \label{wig3} 
\end{multline}
\begin{multline}
(\psi _{0}-\psi _{1})\otimes \phi _{0}\longrightarrow 
 \psi _{0}^{\prime \prime }\otimes \phi _{0}^{\prime \prime }+\psi
_{0}^{\prime \prime }\otimes \phi _{1}^{\prime \prime }+\psi _{1}^{\prime
\prime }\otimes \phi _{0}^{\prime \prime }+\psi _{1}^{\prime \prime }\otimes
\phi _{1}^{\prime \prime },    \label{wig3'}
\end{multline}
Indeed, the conservation law provides an even stronger restriction, and the
last term on the right hand side of (\ref{wig3}) must in fact be zero, and thus at least one of $\psi _{1}^{\prime }$ and $\phi
_{1}^{\prime }$ must always vanish. The same argument applies to \eqref{wig3'} and so (independently), at least one of $\psi
_{1}^{\prime \prime }$ and $\phi _{1}^{\prime \prime }$ must vanish too. 

It follows from (\ref{wig3}) and consistency with the conservation law that  $\psi _{0}^{\prime }$ and $\phi _{0}^{\prime
} $ are necessarily non-zero. For if either did vanish, the right hand side would contain one unit of the 
conserved quantity with certainty, and the left hand side only with probability $\frac{1}{2}$. The same argument runs in clear analogy
for the double-primed quantities. There are then four scenarios that require
consideration:\\

%\begin{case}
%\label{case1}

\noindent {\em Case 1:}  
$\psi _{1}^{\prime }\neq 0$, $\phi _{1}^{\prime }=0$, $\psi
_{1}^{\prime \prime }\neq 0$, $\phi _{1}^{\prime \prime }=0$; 
%\end{case}

%\begin{case}
%\label{case2} 
\noindent{\em Case 2:}
$\psi _{1}^{\prime }\neq 0$, $\phi _{1}^{\prime }=0$, $\phi
_{1}^{\prime \prime }\neq 0$, $\psi _{1}^{\prime \prime }=0$;
%\end{case}

%\begin{case}
%\label{case3} 
\noindent{\em Case 3:}
$\phi _{1}^{\prime }\neq 0$, $\psi _{1}^{\prime }=0$, $\psi
_{1}^{\prime \prime }\neq 0$, $\phi _{1}^{\prime \prime }=0$;
%\end{case}

%\begin{case}
%\label{case4} 
\noindent{\em Case 4:}
$\phi _{1}^{\prime }\neq 0$, $\psi _{1}^{\prime }=0$, $\phi
_{1}^{\prime \prime }\neq 0$, $\psi _{1}^{\prime \prime }=0$.\\
%\end{case}

With this in mind, one can add (\ref{wig3}) and (\ref{wig3'}) to give the
evolution of $\psi _{0}\otimes \phi _{0} $: 
\begin{eqnarray}
2\psi _{0}\otimes \phi _{0} &\longrightarrow &\psi _{0}^{\prime }\otimes \phi
_{0}^{\prime }+\psi _{0}^{\prime }\otimes \phi _{1}^{\prime }+\psi
_{1}^{\prime }\otimes \phi _{0}^{\prime }+  \label{wig4} \\
&&\psi _{0}^{\prime \prime }\otimes \phi _{0}^{\prime \prime }+\psi
_{0}^{\prime \prime }\otimes \phi _{1}^{\prime \prime }+\psi _{1}^{\prime
\prime }\otimes \phi _{0}^{\prime \prime },  \notag
\end{eqnarray}%
and for the evolution of $\psi _{1}\otimes \phi $ we subtract:%
\begin{eqnarray}
2\psi _{1}\otimes \phi _{0} &\longrightarrow &\psi _{0}^{\prime }\otimes \phi
_{0}^{\prime }+\psi _{0}^{\prime }\otimes \phi _{1}^{\prime }+\psi
_{1}^{\prime }\otimes \phi _{0}^{\prime }-  \label{wig7} \\
&&\psi _{0}^{\prime \prime }\otimes \phi _{0}^{\prime \prime }-\psi
_{0}^{\prime \prime }\otimes \phi _{1}^{\prime \prime }-\psi _{1}^{\prime
\prime }\otimes \phi _{0}^{\prime \prime }.  \notag
\end{eqnarray}%
We first consider Case 1 where (\ref{wig4}) and (\ref{wig7})
reduce to 
\begin{multline}
2\psi _{0}\otimes \phi _{0} \longrightarrow 
 \psi _{0}^{\prime }\otimes \phi _{0}^{\prime }+\psi _{1}^{\prime }\otimes
\phi _{0}^{\prime }+\psi _{0}^{\prime \prime }\otimes \phi _{0}^{\prime
\prime }+\psi _{1}^{^{\prime \prime }}\otimes \phi _{0}^{\prime \prime } , 
 \label{wig5'} 
\end{multline}
and
\begin{multline}
2\psi _{1}\otimes \phi _{0} \longrightarrow 
 \psi _{0}^{\prime }\otimes \phi _{0}^{\prime }+\psi _{1}^{\prime }\otimes
\phi _{0}^{\prime }-\psi _{0}^{\prime \prime }\otimes \phi _{0}^{\prime
\prime }-\psi _{1}^{\prime \prime }\otimes \phi _{0}^{\prime \prime }\label{wig34} . 
\end{multline}
Since the left hand side of \eqref{wig5'} contains no units of the conserved quantity, so
must the right, and therefore $\psi _{1}^{\prime }\otimes \phi _{0}^{\prime
}=-\psi _{1}^{^{\prime \prime }}\otimes \phi _{0}^{^{\prime \prime }}$.
Similarly in (\ref{wig34}) the left hand side contains one unit, and if the
right hand side is to agree, we require that $\psi _{0}^{\prime
}\otimes \phi _{0}^{\prime }=\psi _{0}^{\prime \prime }\otimes \phi
_{0}^{\prime \prime }$. \ 

With $\psi _{1}^{\prime }\otimes \phi _{0}^{\prime
}=-\psi _{1}^{^{\prime \prime }}\otimes \phi _{0}^{^{\prime \prime }}$ we
get:
\begin{equation}
2\psi _{0}\otimes \phi _{0} \longrightarrow \psi _{0}^{\prime }\otimes \phi
_{0}^{\prime }+\psi _{0}^{\prime \prime }\otimes \phi _{0}^{\prime \prime },
\end{equation}%
and thus, with $\psi _{0}^{\prime }\otimes \phi _{0}^{\prime }=\psi
_{0}^{\prime \prime }\otimes \phi _{0}^{\prime \prime }$, 
\begin{equation}
\psi _{0}\otimes \phi _{0} \longrightarrow \psi _{0}^{\prime }\otimes \phi
_{0}^{\prime }.
\end{equation}%
Also,%
\begin{equation}
2\psi _{1}\otimes \phi _{0} \longrightarrow \psi _{1}^{\prime }\otimes \phi
_{0}^{\prime }-\psi _{1}^{\prime \prime }\otimes \phi _{0}^{\prime \prime },
\label{wig14}
\end{equation}%
and finally, exploiting the condition $\psi _{1}^{\prime }\otimes \phi _{0}^{\prime
}=-\psi _{1}^{\prime \prime }\otimes \phi _{0}^{\prime \prime }$, we arrive
at
\begin{equation}
\psi _{1}\otimes \phi _{0} \longrightarrow \psi _{1}^{\prime }\otimes \phi
_{0}^{\prime }.  \label{wig17}
\end{equation}

We now consider Case 2 which, with \eqref{wig4} gives
\begin{equation}
2\psi _{0}\otimes \phi _{0} \longrightarrow \psi _{0}^{\prime }\otimes \phi
_{0}^{\prime }+\psi _{1}^{\prime }\otimes \phi _{0}^{\prime }+\psi
_{0}^{\prime \prime }\otimes \phi _{0}^{\prime \prime }+\psi _{0}^{\prime
\prime }\otimes \phi _{1}^{\prime \prime },
\end{equation}%
and thus one might wish to conclude that $\psi _{1}^{\prime }\otimes \phi
_{0}^{\prime }=-\psi _{0}^{\prime \prime }\otimes \phi _{1}^{\prime \prime }$%
. However, this can never be satisfied; these vectors must be distinct
unless they are both zero (which is excluded, by assumption), since the 
unit of conserved quantity resides in different Hilbert spaces. 

Case 3 gives 
\begin{equation}
2\psi _{0}\otimes \phi _{0} \longrightarrow \psi _{0}^{\prime }\otimes \phi
_{0}^{\prime }+\psi _{0}^{\prime }\otimes \phi _{1}^{\prime }+\psi
_{0}^{\prime \prime }\otimes \phi _{0}^{\prime \prime }+\psi _{1}^{\prime
\prime }\otimes \phi _{0}^{\prime \prime }  \label{wig6}
\end{equation}%
and we conclude that it must be the case that $\psi _{0}^{\prime }\otimes
\phi _{1}^{\prime }=-\psi _{1}^{\prime \prime }\otimes \phi _{0}^{\prime
\prime }$ which, again, cannot be fulfilled for both non-zero. We therefore must also
reject Case 3.

Finally Case 4 gives
\begin{equation}
2\psi _{0}\otimes \phi _{0} \longrightarrow \psi _{0}^{\prime }\otimes \phi
_{0}^{\prime }+\psi _{0}^{\prime }\otimes \phi _{1}^{\prime }+\psi
_{0}^{\prime \prime }\otimes \phi _{0}^{\prime \prime }+\psi _{0}^{\prime
\prime }\otimes \phi _{1}^{\prime \prime }
\end{equation}
and
\begin{equation}
2\psi _{1}\otimes \phi _{0} \longrightarrow \psi _{0}^{\prime }\otimes \phi
_{0}^{\prime }+\psi _{0}^{\prime }\otimes \phi _{1}^{\prime }-\psi
_{0}^{\prime \prime }\otimes \phi _{0}^{\prime \prime }-\psi _{0}^{\prime
\prime }\otimes \phi _{1}^{\prime \prime },
\end{equation}
and so $\psi _{0}^{\prime }\otimes \phi _{1}^{\prime }=-\psi _{0}^{\prime
\prime }\otimes \phi _{1}^{\prime \prime }$ and $\psi _{0}^{\prime
}\otimes \phi _{0}^{\prime }=\psi _{0}^{\prime \prime }\otimes \phi
_{0}^{\prime \prime }$. 

It is now evident that each of the permissible cases
gives the same state evolution for $\psi _{0}\otimes \phi $; Case 4 yields
\begin{equation}
2\psi _{0}\otimes \phi _{0} \longrightarrow \psi _{0}^{\prime }\otimes \phi
_{0}^{\prime }+\psi _{0}^{\prime \prime }\otimes \phi _{0}^{\prime \prime },
\label{wig8}
\end{equation}%
and with $\psi _{0}^{\prime }\otimes \phi _{0}^{\prime }=\psi _{0}^{\prime
\prime }\otimes \phi _{0}^{\prime \prime }$, we arrive at

\begin{equation}
\psi _{0}\otimes \phi _{0} \longrightarrow \psi _{0}^{\prime }\otimes \phi
_{0}^{\prime }.  \label{wig18}
\end{equation}%
However, for the evolution of $\psi _{1}\otimes \phi $, using $\psi _{0}^{\prime }\otimes \phi _{1}^{\prime }=-\psi
_{0}^{\prime \prime }\otimes \phi _{1}^{\prime \prime }$, we see that a different form emerges than from Case 1:
\begin{equation}
\psi _{1}\otimes \phi _{0} \longrightarrow \psi _{0}^{\prime }\otimes \phi
_{1}^{\prime }.
\end{equation}

With these considerations, we now summarise the possible forms of the evolution of $(\psi _{0}+\psi
_{1})\otimes \phi _{0} $ and $(\psi _{0}-\psi _{1})\otimes \phi _{0} $.  Remembering
that the only cases which contain, \textit{a priori}, no contradiction, are Cases 1 and 4,
the first scenario is that Case 1 is satisfied, and we have:
\begin{equation}
(\psi _{0}+\psi _{1})\otimes \phi _{0} \longrightarrow (\psi _{0}^{\prime }+\psi
_{1}^{\prime })\otimes \phi _{0}^{\prime }  \label{wig15},
\end{equation}%
and
\begin{equation}
(\psi _{0}-\psi _{1})\otimes \phi _{0} \longrightarrow (\psi _{0}^{\prime }-\psi
_{1}^{\prime })\otimes \phi _{0}^{\prime }.  \label{wig16}
\end{equation}%
This cannot represent a measurement in any ordinary or physically
meaningful sense, since the final states of the apparatus coincide on the
right hand side of (\ref{wig15}) and (\ref{wig16}), leaving us in the
position that there is no way of distinguishing which eigenstate of $S_{x}$
had been present on the left hand side. Furthermore, this product form
does not correspond to a modification of equations (\ref{intro1}) and (\ref
{intro2}) (as is claimed by Wigner).

\bigskip The second scenario is that Case 4
is satisfied, and we see that summing (\ref{wig17}) with (\ref{wig18}) gives:
\begin{equation}
(\psi _{0}+\psi _{1})\otimes \phi _{0} \longrightarrow \psi _{0}^{\prime }\otimes
(\phi _{0}^{\prime }+\phi _{1}^{\prime })  \label{wig12}
\end{equation}%
and subtracting: 
\begin{equation}
(\psi _{0}-\psi _{1})\otimes \phi _{0} \longrightarrow \psi _{0}^{\prime }\otimes
(\phi _{0}^{\prime }-\phi _{1}^{\prime })  \label{wig13}
\end{equation} 
This coincides with \eqref{mmt1} and \eqref{mmt2} (Sec. \ref{subsec:alternative}), 
and is the same result as Wigner obtained on his final page.

%\bigskip

\end{document}